# Automatic Expansion of Domain-Specific Affective Models for Web Intelligence Applications

Albert Weichselbraun[1,4] · Jakob Steixner[2] · Adrian M.P. Braşoveanu[2] · Arno Scharl[3,4] · Max Göbel[4] · Lyndon J. B. Nixon[2,3]

## Abstract

Sentic computing relies on well-defined affective models of different complexity—polarity to distinguish positive and negative sentiment, for example, or more nuanced models to capture expressions of human emotions. When used to measure communication success, even the most granular affective model combined with sophisticated machine learning approaches may not fully capture an organisation's strategic positioning goals. Such goals often deviate from the assumptions of standardised affective models. While certain emotions such as *Joy* and *Trust* typically represent desirable brand associations, specific communication goals formulated by marketing professionals often go beyond such standard dimensions. For instance, the brand manager of a television show may consider *fear* or *sadness* to be desired emotions for its audience. This article introduces expansion techniques for affective models, combining common and commonsense knowledge available in knowledge graphs with language models and affective reasoning, improving coverage and consistency as well as supporting domain-specific interpretations of emotions. An extensive evaluation compares the performance of different expansion techniques: (i) a quantitative evaluation based on the revisited *Hourglass of Emotions* model to assess performance on complex models that cover multiple affective categories, using manually compiled gold standard data, and (ii) a qualitative evaluation of a domain-specific affective model for television programme brands. The results of these evaluations demonstrate that the introduced techniques support a variety of embeddings and pre-trained models. The paper concludes with a discussion on applying this approach to other scenarios where affective model resources are scarce.

**Keywords** Affective models · Hourglass of emotions · Language models · Embeddings · Knowledge graphs

✉ Albert Weichselbraun
albert.weichselbraun@fhgr.ch;
weichselbraun@weblyzard.com

Jakob Steixner
steixner@modultech.eu

Adrian M.P. Braşoveanu
adrian.brasoveanu@modul.ac.at; brasoveanu@modultech.eu

Arno Scharl
arno.scharl@modul.ac.at; scharl@weblyzard.com

Max Göbel
goebel@weblyzard.com

Lyndon J. B. Nixon
nixon@modultech.eu; lyndon.nixon@modul.ac.at

1    University of Applied Sciences of the Grisons, Chur, Switzerland

2    MODUL Technology, Vienna, Austria

3    MODUL University Vienna, Vienna, Austria

4    webLyzard technology, Vienna, Austria

## Introduction

Organisations use Web intelligence applications to obtain real-time insights into the public perception of their brands. Driven by news media coverage, influential social media postings and real-world events, the mood of consumers and their perception of a brand can change rapidly. Consumers who discuss brands through digital channels not only respond to communication, but also play a pivotal role in shaping brand reputation, for example when repeating or commenting on a story. This reflects their personal connection to the brand and the collective nature of brand authorship. *Sentiment* and *affective categories* are important indicators derived from these user actions. They help organisations to better understand the public debate and track evolving perceptions of their brands. To measure communication success, however, general emotional categories often do not suffice. Domain-specific affective models incorporate specific emotional categories that are not found in general models. Their interpretation might also deviate from typical perceptions. (In the case of a television show, for example, *fear* or *sadness* may represent desirable associations.) Another advantage of domain-specific affective models is the possibility to include not only emotional categories but also other desired or undesired semantic associations specific to the situational context.

To provide actionable knowledge for communication professionals, Web intelligence applications should therefore support both: (i) standardised affective models to benchmark multiple brands and compare the results with third-party studies and (ii) domain-specific affective models that consider the specific communication goals of an organisation. Often formulated in an ad hoc manner, e.g. during a communications workshop, the major challenge of such models relates to their inconsistent and often incomplete nature. They tend to have low coverage since the time and effort invested into their definition and disambiguation cannot compete with standardised affective models based on many years of scientific research.

Addressing this problem, this article introduces a data-driven method to expand domain-specific affective models in situations when lexical resources required as training data are scarce. The method automatically extends such models through knowledge graph concepts in conjunction with language models and affective reasoning. The goal is to improve the coverage and consistency of affective models such as the *webLyzard Stakeholder Dialogue and Opinion Model* (WYSDOM),[1] which provides a communication success metric that combines sentiment and emotional categories with desired and undesired semantic associations. Computed based on co-occurrence patterns, these associations provide real-time insights into the success of marketing and public outreach activities.

WYSDOM goes beyond sentiment and standardised emotional categories by asking communication experts to specify the intended positioning of their organisations. This positioning is expressed in the form of desired and undesired keywords. In the case of the U.S. National Oceanic and Atmospheric Administration (NOAA), for example, an association with "climate change" in the public debate indicates successful communication although the term typically carries a negative sentiment [1].

Tracking the WYSDOM metric over time allows to assess to what extent a chosen communication strategy impacts the public debate, how consistently a message is being conveyed and whether this message helps to reinforce the intended brand positioning in a sustainable manner. The visual representation of the metric comes is a stacked bar chart that combines content-based metrics (positive vs. negative sentiment and desired vs. undesired associations) with other indicators of success such as page views and the number of visits (see the lower half of Fig. 1). While the metric has initially been created for tracking the success of brand communication, it is applicable to a wide range of use cases that benefit from a hybrid display of content-based metrics in conjunction with other *Key Performance Indicators* such as sales figures or stock market prices.

We have investigated several affective models for possible inclusion of their emotional categories into the WYSDOM metric, as shown in Fig. 1. These models include *Sentiment* (Positive, Neutral, Negative), *Brand Personality* (Sincerity, Competence, Ruggedness, Sophistication, Excitement and the added dimension Sustainability) according to Aaker [2], Plutchik's *Wheel of Emotions* [3], as well as Cambria et al.'s *Hourglass of Emotions* [4]. The latter provides a comprehensive multidimensional framework for interpreting emotions used in diverse fields such as ontology construction [5] and affective visualisation [6].

This paper showcases a method to augment domain-specific affective models with concepts extracted from open knowledge graphs such as ConceptNet [7] and WordNet [8], lexical resources, pre-trained embeddings like Global Vectors (GloVe) [9], domain documents from multiple sources and language models such as the Bidirectional Encoder Representations from Transformers (BERT) and its distilled version known as DistilBERT [10]. It provides a fast and reliable method to build domain-specific affective classifiers even if resources required for the tasks are scarce.

The rest of the article is organised as follows: Section 2 and Section 3 provide an overview of affective models and related work. Section 4 introduces the affective model expansion method, as well as the affective knowledge extraction method used for the evaluations. Section 5 presents the gold standard (Section 5.1) and discusses two use cases: (i) a quantitative evaluation that expands an affective model based on the revised Hourglass of Emotions (Section 5.2) and (ii) a qualitative evaluation that demonstrates the method's suitability for improving domain-specific affective models (Section 5.3). Section 6 discusses the evaluation results. The concluding Section 7 summarises the contribution and highlights the strengths and weaknesses of the presented method.

## Affective Models

This section first provides background information on sentiment and emotion classification models. It then outlines the relation of these models to work on domain-specific affective models.

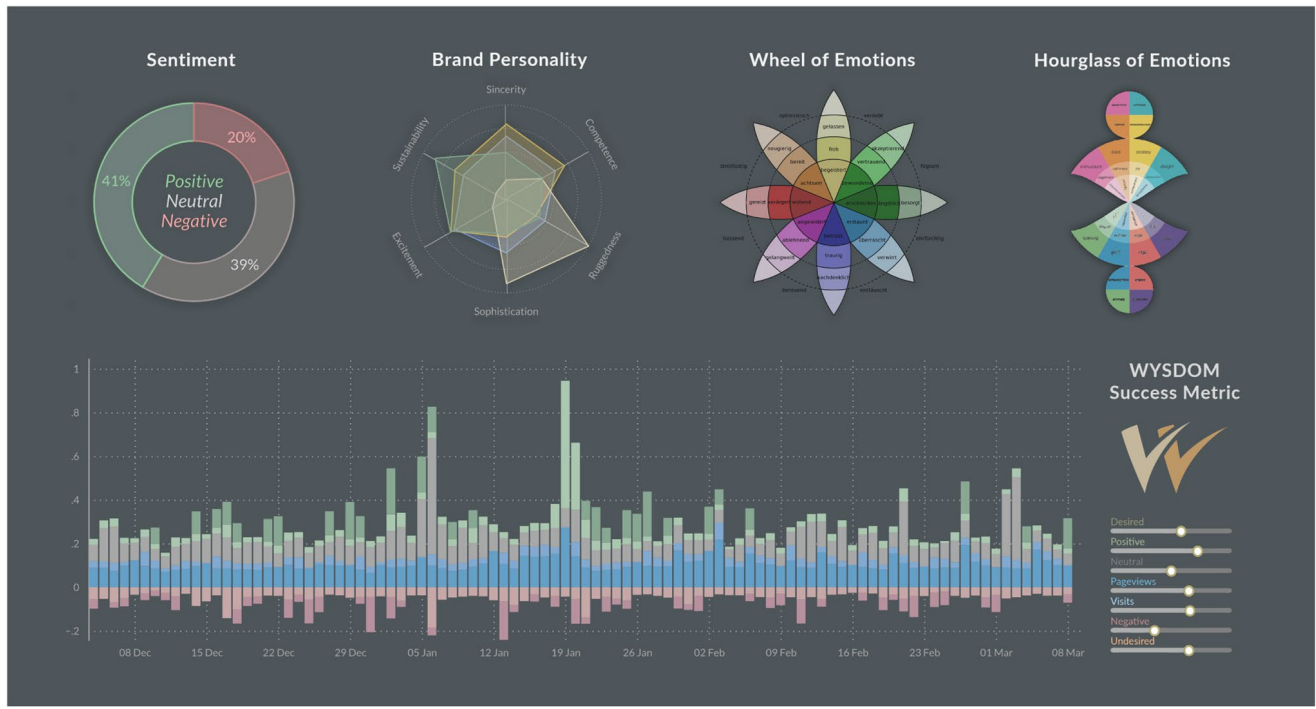

**Fig. 1** Overview of affective models including *Sentiment, Brand Personality*, *Wheel of Emotions*, *Hourglass of Emotions* and the *webLyzard Stakeholder Dialogue and Opinion Model* (WYSDOM)

## Sentiment and Emotion Classification Models

Sentiment analysis aims at determining whether a statement is positive (e.g. *'Awesome battery life.'*), negative (e.g. *'Do not waste your money on this phone!'*), neutral (e.g. *'I bought this for my spouse.'*) or ambivalent (e.g. mixes multiple polarities, *'Great display but horrible battery life.'*). Emotion analysis, in contrast, provides a much more fine-grained classification by recognising the emotion(s) expressed in a text and mapping them to emotional categories.

Research in this area has significantly gained traction in recent years, developing classification models that draw on psychology [11, 12], neuroscience [12], social science [13], computer science [14] and engineering [15, 16]. The complexity of emotion processing has led to many definitions and interpretations that differ in the specific aspects considered: physiological processes, evolutionary adaptation to environmental stimuli, affective evaluation or the subjectivity of emotional experience [17]. This variety is reflected in an extensive review of affective models and algorithms by Wang et al. [18], which covers nine popular models and 65 emotions discussed in these models.

Many authors have formulated classification systems of emotions such as Ekman's basic emotions [19]. Some are well-known in business and marketing, such as the Wheel of Emotions [3], the Circumplex Model of Affect [20] or the Hourglass of Emotions [4] and its revised version [12]. The structure of these frameworks often distinguishes basic and derived emotions, e.g. *envy* derived from the combination of *shame* and *anger*.

The *Hourglass of Emotions* is a comprehensive and multidimensional framework for interpreting emotions, inspired by neuroscience and motivated in psychology. The initial version [4] distinguished four affective categories: pleasantness (defined as *joy–sadness*, based on the affective concepts in this category), attention (*anticipation–surprise*), sensitivity (*anger–fear*) and aptitude (*trust–disgust*). By using the different activation scales (e.g. *pleasantness* can have different activation levels characterised as *ecstasy*, *joy*, *serenity*, *pensiveness*, *sadness* and *grief*) and composition, the model can express a wide range of emotions. The Hourglass of Emotions has also been used extensively in information visualisation due to its colour associations. Recently, a revised version [12] was published that further improved the original model by removing neutral emotions, increasing consistency (e.g. comfort and discomfort are now classified as opposites), and adding polar emotions as well as self-conscious emotions. The model also refined the colour scheme to be in line with recent studies on colour-emotion associations and considerably improved the polarity scores obtained for compound emotions. The revised model changes the definition of the four primary affective categories and the associated affective concepts into introspection (*joy–sadness*), temper

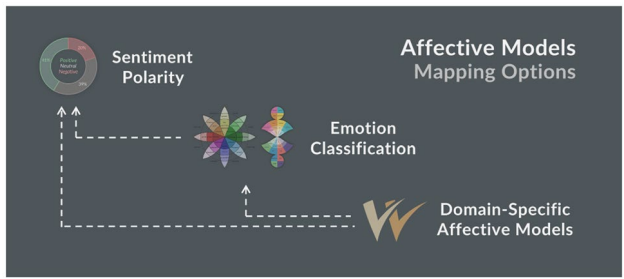

**Fig. 2** Types of affective models including possible mappings, as indicated by the dotted lines

(*calmness–anger*), attitude (*pleasantness–disgust*) and sensitivity (*eagerness–fear*), while also providing additional explanations on how to create compound emotions using the new emotion classification scheme. Some researchers might consider the elimination of *surprise* problematic, but in terms of classification it is a welcome improvement since most annotators found it difficult to decide whether it should be considered a positive or negative emotion.

## Domain-Specific Affective Models

Sentiment polarity and emotion categories often do not coincide with desired business outcomes such as the brand perception that is aspired for or an organisation's public relations goals. Scharl el al. [1] have therefore introduced the WYSDOM success metric that allows companies to define domain-specific affective models. These models aim at capturing affective content relevant to their specific business communication goals.

We consider "affective models" as an umbrella term that covers sentiment polarity, standard affective models using common emotion categorisations as well as domain-specific affective models. Each model covers different affective dimensions (e.g. sentiment polarity or the emotions defined by the specific affective model) and might even provide mappings to and from other models. SenticNet 6, for instance, translates text into primitives and subsequently superprimitives from which it inherits a specific set of emotions that in turn can be mapped to a particular polarity [21].

Figure 2 outlines the relation between three types of affective models: (i) *Sentiment polarity* is the best understood model from a research perspective. It only includes one affective dimension to distinguish positive, neutral and negative documents. (ii) *Emotion classification* models that consider multiple affective dimensions are more challenging, which is reflected in the large number of models from different authors. (iii) *Domain-specific affective models* follow a more customised approach, relating affective content to the specific communication goals of an organisation. They typically focus on a small number of affective dimensions, given their customised nature and the manual effort involved in creating the model's specifications and continuously updating them in line with evolving communication goals. The dotted lines in the figure indicate possible mappings between these models, for

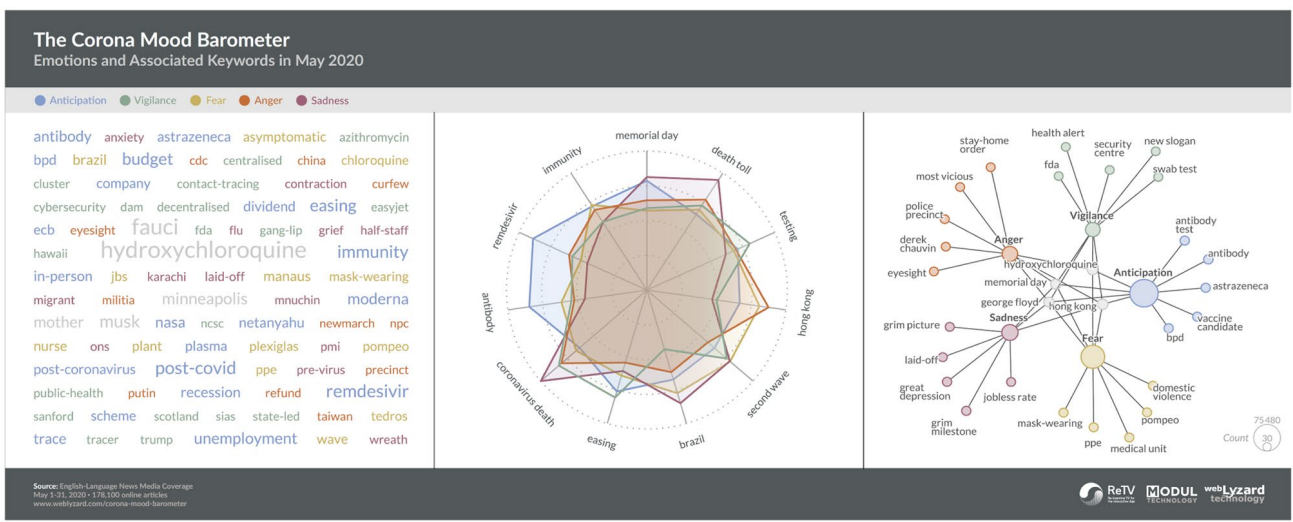

**Fig. 3** Affective analysis of the media coverage on the COVID-19 pandemic based on Plutchik's *Wheel of Emotions*, using a tag cloud (left), a radar chart (middle) and keyword graph (right) with colour coding to distinguish selected emotions including *Anticipation, Vigilance, Fear, Anger* and *Sadness*

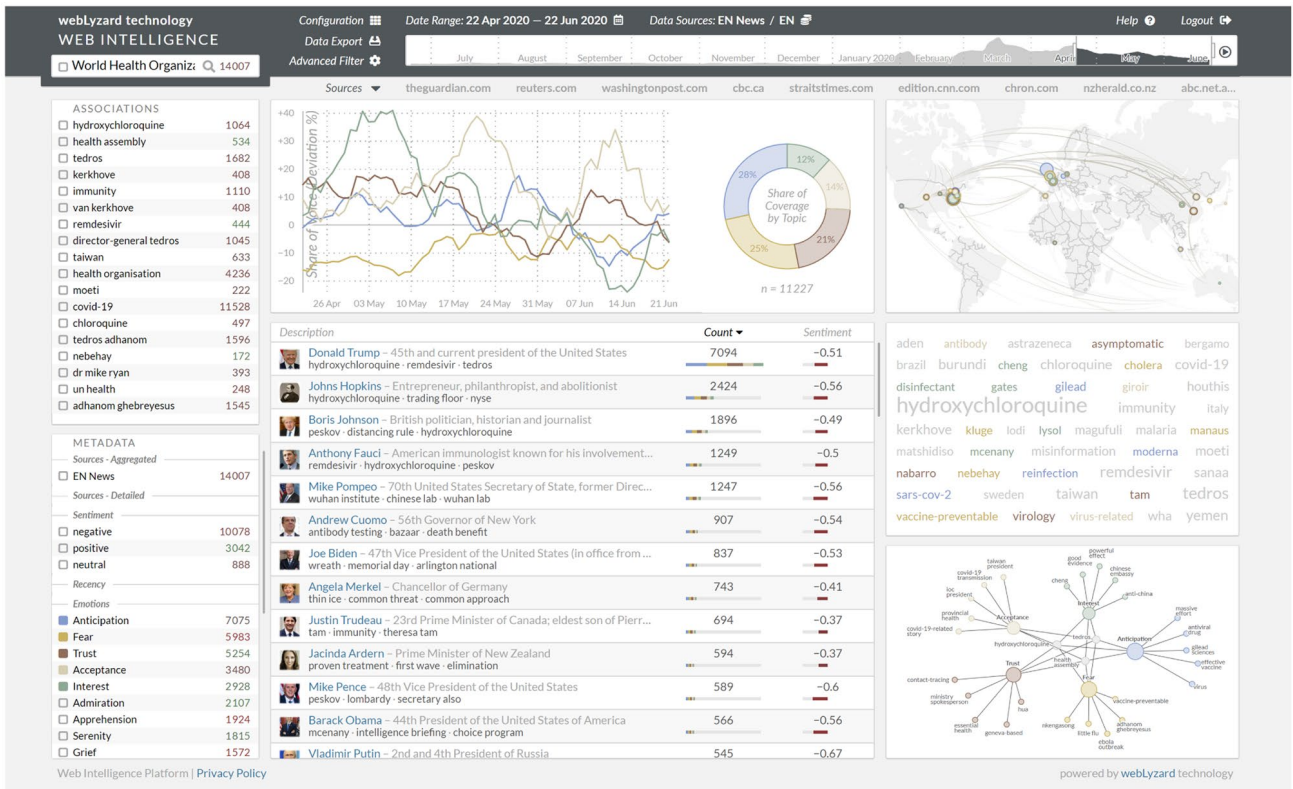

**Fig. 4** Screenshot of the webLyzard Web intelligence dashboard with results of a query on *World Health Organisation* between April and June 2020—using colour coding to visualise the emotions *Anticipation, Fear, Trust, Acceptance* and *Interest*

example from emotions to sentiment polarity as outlined by Cambria et al. [12].

Figure 3 shows a previous application of affective analysis conducted as part of the *Corona Mood Barometer*.[2] The system uses a combination of story detection and emotion analysis techniques to discover what drives the public coronavirus 2019 (COVID-19) debate and how government responses to the coronavirus pandemic are perceived across the various countries. The visualisations explore associations with five selected emotions (Anticipation, Vigilance, Fear, Anger and Sadness) to better understand the drivers of the public debate in May 2020. The tag cloud sorts associations alphabetically, colour coding them by emotion. The radar chart projects the top keywords along multiple axes, revealing the relative strength of association with each emotional category. The keyword graph then applies a hierarchical layout, with grey centre nodes to represent keywords linked to multiple emotional categories.

Figure 4 shows how the visualisations can be accessed via the webLyzard Web intelligence dashboard, based on a search query for the *World Health Organization* (WHO) that resulted in more than 14,000 documents published between April and June 2020. The dashboard is an advanced information exploration and retrieval interface that helps to track the various emotions along multiple context dimensions (sources, regions, languages, etc.) and enables on-the-fly filtering and query refinement options to access a comprehensive content repository of news and social media content.

## Related Work

The discussion of related work starts with deep learning techniques for classifying affective categories such as emotions, then provides an overview of methods used for extracting content-based communication success metrics from news and social media, and a discussion of how Natural Language Processing (NLP) can support this process. Additional background information on this area of research is available in recent surveys such as those published in Cambria et al. [22], Xing et al. [23], Chaturvedi et al. [24] and Mehta et al. [25].

---

[2] www.weblyzard.com/corona-mood-barometer

## Deep Learning for Affective Classification

During the last decade, the paradigm of affective models shifted due to the rise of language models based on deep learning methods that capture the meaning of all the words from a text corpus as a set of vectors in a lower-dimensional space (e.g. embeddings like word2vec [26], GloVe [9], and fastText [27]). With the addition of attention mechanisms and Transformer architectures [28], current language models such as BERT [29] were shown to be better at picking up linguistic phenomena [30] and performing tasks as diverse as capturing analogies, semantic role labelling, textual entailment, sentiment analysis and named entity recognition (NER). NER is particularly important for processes like coreference and anaphora resolution [31]. Coreference resolution refers to the process of finding all expressions that refer to the same object or entity and linking them to a single identifier, whereas anaphora resolution is the process through which the antecedent of an expression is determined. These tasks are important for affective reasoning as they support subjectivity detection [24], which helps us understand whether a text refers to the subject or is the subject's opinion on a product or review. *Subjectivity detection* [24] is a complex problem that has moved through various stages from manually crafted features and bootstrapping to syntactic features and domain adaptation via knowledge graphs and neural networks and finally to cross-modal fusion of text, video and audio via Transformer models like BERT. While hand-crafted features produced many false positives, classic machine learning (ML) with syntactic features missed even shallow representations of meaning, which were later added by knowledge graphs. Combining subjectivity detection with advanced filtering mechanisms such as threat or sarcasm detection [32], cause-pairs extraction [33] or concept-level sentiment extraction [34] enables a fine-grained approach towards affective classification and provides support for operations like removal of bias or propaganda, provision of better context awareness, and better accuracy of subjectivity values. The next trend in subjectivity detection research seems to be focused on distinguishing cultural aspects, biases, nuances and dialects.

Affective classification is usually framed as a text classification task, as outlined by Kowsari et al. [35] and Wolf et al. [10] who surveyed deep learning architectures used for sentiment analysis. A multi-layer perceptron (MLP) stacked ensemble is used for predicting emotional intensity for different content types such as Twitter postings, microblogs and news [36]. It showed significant improvements over similar systems for both generic emotion analysis and financial sentiment analysis. The EvoMSA [37] open-source multilingual toolkit for creating sentiment classifiers composes the outputs of multiple models (fastText, Emoji Space, lexicon-based model, etc.) into a vector space that is then wired into the EvoDAG genetic algorithm to predict the final class. The performance improvements brought by EvoMSA are impressive, but the resulting models obtained from the EvoDAG algorithm were not designed to be explainable.

The key element of successful NLP language models is a process called knowledge transfer, which refers to the transfer of learnt patterns (e.g. weights) from one problem to another [38]. In some cases, if there is a need to reduce the size of the models and a small reduction in performance is acceptable, it is also possible to use a process called knowledge distillation [10], a lightweight knowledge transfer process for compressed models that are smaller and faster. Due to the knowledge transfer process, Transformer models such as BERT pick up various linguistic phenomena like direct objects, noun modifiers and coreferents [30], which benefits their performance in tasks like sentiment analysis and NER [29].

Another key research problem is the adaptation of affective models to new domains. This is necessary since many domains introduce special terminology or jargon, which can lead to misinterpretations of the affective categories they convey. In contrast to the work introduced in this paper, domain adaptation techniques do not create domain-specific affective models but rather fine-tune existing models (e.g. sentiment polarity). Xing et al. [39] present a cognitive-inspired adaptation method that emulates metacognition processes for detecting contradictions and obtaining the correct sentiment polarity of words when a human is confronted with a new language domain. Murtadha et al. [40] use weak supervised methods to cluster words according to their sentiment polarity for aspect-based sentiment analysis in the target domains. They also apply an attention-based long short-term memory (LSTM) network to the same task. Both models work well due to the weight reduction for non-sentiment parts from a sentence. Zhao et al. [41] discuss multi-source domain adaptation for a cross-domain sentiment classification task. Their paper shows that joint learning for cross-domain tasks leads to good results and a greater generalisation capability, while at the same time enabling deep domain fusion. Domain adaptation techniques can also vary depending on the architecture. Since Transformers like BERT are generally task-agnostic, the best method to adapt them to another domain is optimising training by applying strategies such as adversarial training, pre-training and post-training [42].

## Extraction of Content-Based Communication Success Metrics from Web and Social Media

The extraction of communication success metrics from digital content streams is a dynamic research area that relies heavily on NLP and information visualisation. Traditionally, sentiment is among the most frequently used metrics for evaluating the impact of a campaign. The required computation can therefore be formulated as sentiment polarity extraction (e.g. the identification or positive or negative emotions) or stance classification (e.g. the classification of opinions towards a certain target) [42]. The task of determining the polarity of the considered resources typically leverages lexical resources that map numerical values to terms with an affective meaning (e.g. the value -1 for terms on the negative affective scale such as *fear*, or +1 for positive terms such as *trust*). Lexical resources for this purpose show different levels of granularity, e.g. mere polarity vs. subjectivity vs. fine-grained aspects.

Related analytic tasks focus on event discovery and the representation of relations as a knowledge graph, as presented by Nguyen et al. [43] and Camacho et al. [44]. The integration of knowledge graphs with ML approaches aids the recognition of emotions in languages such as Arabic and Spanish for a wide variety of NLP tasks such as knowledge transfer and machine translation [44]. Sentic computing [45] helps to model categories such as life satisfaction and safety, because it provides linguistic cues on emotions such as *sadness, joy, anger* and *fear*. Regardless of the computational methods used, the goal of the sentiment analysis is to produce values stemming from differences between evaluative ratings of positive and negative emotions.

Well-known ontologies for sentiment analysis include OntoSenticNet [5] and the multilingual visual sentiment ontology [46]. OntoSenticNet [5] acts as a commonsense ontology for the sentiment domain. The visual sentiment ontology [46] is used in the multimedia domain. Since such ontologies often cover a limited number of domains, a semi-automatic ontology builder was proposed for solving the issue of domain adaptation for aspect-based sentiment analysis [47]. In addition to ontologies, knowledge graphs are used to support complementary tasks like entity detection and commonsense reasoning. DBpedia [48] and Wikidata [49] are public knowledge graphs typically used in entity linking tasks, whereas ConceptNet [7] and SenticNet [21] are used in sentiment and emotion detection. The last version of SenticNet used subsymbolic artificial intelligence (e.g. clustering, recommendation algorithms) to detect patterns in natural language and represent them with symbolic logic in a knowledge graph. The key to understanding the construction of the latest version of SenticNet is the idea of language composition, namely the fact that multi-word expression can be deconstructed into primitives or superprimitives (e.g. functions that will be able to represent an entire range of primitives). To compute the polarity, it suffices to look up the value of the superprimitives associated with the respective primitive.

Many papers are dedicated to the construction of domain-specific affective knowledge graphs for aspect-based analysis. One method to create such graphs is showcased in Cavalleri et al. [50] and is based on graph embeddings, namely the embedding of entire communities instead of individual nodes. This leads to significant improvement in applications like community detection or node classification. Another method for constructing knowledge graphs is presented by Ghosal et al. [51]: the filtering of ConceptNet to create a domain-aggregated graph that is then fed to a graph convolutional network (GCN) autoencoder to build a domain-adversarial training dataset, which includes both domain-specific and domain-agnostic concepts. Du et al. [42] leverage learnt entity and relation embeddings to fully exploit the constraints of a commonsense knowledge graph. Bijari et al. [52] combine sentence-level graph-based learning representations with latent and continuous features extraction to improve sentiment polarity detection. Custom affective graphs are built in order to provide both word sense disambiguation and affective reasoning for specific domains like law and medicine (e.g. K-BERT [53]) and finance (e.g. FOREX market prediction [54]). Finally, a recent survey expands upon the problems of building, representing and applying knowledge graphs for affective reasoning [55].

## Method

Many content-based success metrics consider affective categories like sentiment polarity, emotions and domain-specific metrics such as (un)desired keyword associations, e.g. for the above-mentioned WYSDOM model. They require multi-faceted sentic computing engines that integrate *syntactics* (e.g. part-of-speech tagging, chunking, lemmatisation), *semantics* (e.g. topic extraction, named entities) and *pragmatics* (e.g. sarcasm detection, aspect extraction) layers [22]. The authors have developed components across all these layers, including an aspect-based sentiment analysis engine [56], a named entity linking (NEL) engine [57] and a NLP and visualisation pipeline that includes topic, concept and story detection [58]. These components are used as a basis for the extraction of social indicators. The computed sentiment values typically include aspect, polarity and subjectivity. Aspect is used to analyse the various features of products or ideas, polarity offers the details about sentiment orientation (positive, neutral or negative), and subjectivity describes a person's opinion towards a product, topic or idea.

Developing the method to expand domain-specific affective models has been guided by the following goals and constraints:

1. The method should be applicable in both research and corporate settings, covering different types of affective models (comprehensive standardised affective lexicons as well as tailored domain-specific models such as WYSDOM).
2. The expansion process should not require large and comprehensive corpora, since creating such corpora is not feasible in most industrial settings.
3. The expansion process must perform well across domains and should draw on publicly available resources such as common and commonsense knowledge, pre-trained word embeddings and language mod-els to improve (by disambiguating terms based on their context) and expand (by extending the affective lexicon) affective models.

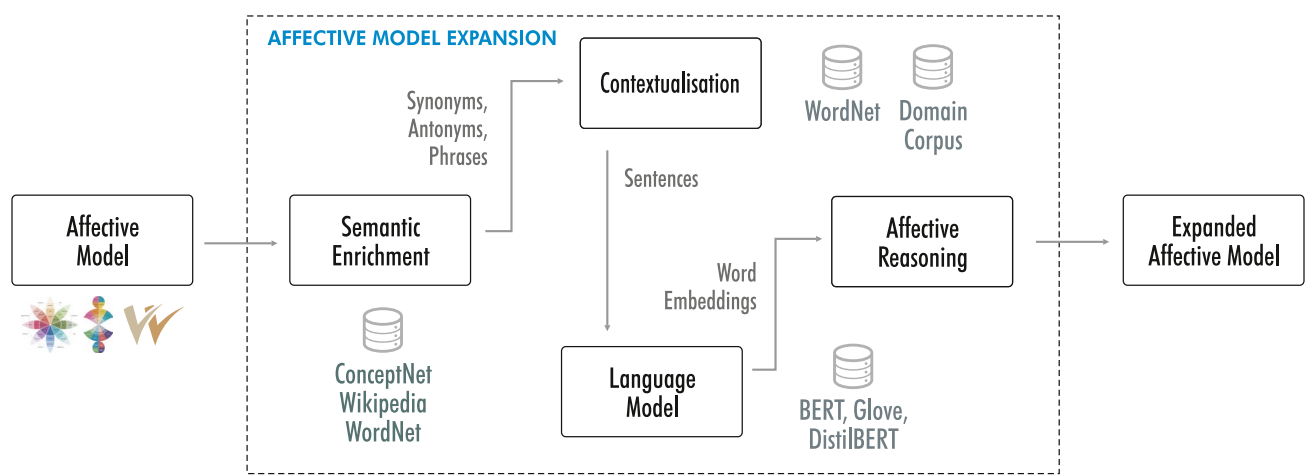

**Fig. 5** Semantic enrichment, contextualisation and affective reasoning based on language models and word embeddings for affective knowledge expansion

4. The expanded affective models should be usable with simple lexicon-based sentiment analysis techniques as well as with more sophisticated approaches that consider syntactical (i.e. negation, modifiers, quotes, etc.) and contextual (i.e. disambiguation of the term based on its actual use in a sentence) information.

Section 4.1 introduces the affective model expansion method. Section 4.2 then presents an affective knowledge extraction technique that builds on the expanded models and considers the sentence's grammar and context in the extraction process.

## Affective Model Expansion

The affective model expansion technique uses explicit knowledge available in lexical databases and knowledge graphs such as WordNet [8], ConceptNet [7] and Wikidata [49] as well as implicit knowledge about a term's semantics encoded in word embeddings and language models. Figure 5 outlines the iterative affective model expansion process in greater detail. The method enriches terms and phrases from the seed model with structured knowledge obtained from publicly available sources such as WordNet, ConceptNet and Wikidata by mining synonyms, antonyms and phrases that are related to the seed terms.

The next step aims at contextualising the enriched model by mining WordNet and domain corpora for sentences that contain concepts from the affective model. Contextualisation does not only facilitate the use of language models such as BERT and DistilBERT, but also allows improving the precision and consistency of the affective model by splitting ambiguous terms into multiple concepts (senses). The example sentences that demonstrate a term's use in a context are transformed into the embedding space and disambiguated based on Algorithm 1, which is part of the affective reasoning component. Table 1 provides examples that illustrate the outcome of this process based on the ambiguous terms *like*, *probe* and *project* with two selected senses for each term. The left side of the table shows the affective categories assigned to the terms' average senses, while the right-hand columns present the affective categories assigned to the disambiguated senses.

---

**Algorithm 1:** Disambiguates a term ($t$) with the affective categories ($s$) by computing a dictionary $s\_dict$ of all the term's senses and the corresponding affective categories.

**Data:** $t, s_t$
**Result:** s_dict[sense]

1 senses ← get_senses(t) ;
2 sense_vector_dict ← {} ;
3 s_dict ← {} ;
   /* compute the centroid of the term's senses and the average distance between senses. */
4 **foreach** *sense in senses* **do**
5    usage_examples ← get_example(sense) ;
6    usage_vectors ← lm(usage_examples) ;
7    sense_vector_dict[sense] ← avg(usage_vectors) ;
8 **end**
9 sense_centroid ← avg(sense_vector_dict) ;
10 avg_sense_dist ← avg_dist(sense_vector_dict) ;
   /* validate affective categories of senses. */
11 **foreach** *sense in senses* **do**
12    **if** *(cos(sense_vector_dict[sense], sense_centroid ≤ avg_sense_dist · 1.3)* **then**
13       | s_dict[sense] ← $s_t$ ;
14    **end**
15    **else**
16       | s_dict[sense] ←
          get_affective_categories(sense) ;
17    **end**
18 **end**
19 **return** *s_dict*;

---

For terms that are available in WordNet, the algorithm loops over all senses, retrieves examples of each sense and uses the language model to transform them into the corresponding embedding space for that particular sense. Algorithm 1 then computes a centroid that represents the term's average usage and computes the senses' average distance from this centroid. Finally, we assign the senses' overall values for all semantic categories to senses that are close to the term's average usage and compute a refined set of semantic categories for terms that are different from the seed term's average usage. For terms not covered in WordNet, Algorithm 1 is modified to use example sentences mined from the domain corpus rather than on explicit WordNet senses.

Once all multi-sense terms have been successfully disambiguated, the affective reasoning component performs a proximity search as outlined in Algorithm 2 to further expand the affective model. The expanded model may then run through another iteration to enrich the extracted concepts, contextualise them and transform all terms into embedding space.

---

**Algorithm 2:** Proximity lookup and affective reasoning heuristic used for computing the semantic categories ($sc$) of a new term ($t$) based on the language model $lm$.

---

**Data:** $t$, $lm$, affective_categories
**Result:** $s$

1  $s \leftarrow \{\}$ ;
2  $pt \leftarrow$ get_proximate_terms($lm$, $t$) ;
3  $ant \leftarrow$ identify_antonyms($lm$, $pt$) ;
4  $sim \leftarrow pt - ant$ ;
5  **foreach** $ac$ in affective_categories **do**
6  $\quad$ $avg\_similar \leftarrow$ avg_affective_value($sim$, $ac$);
7  $\quad$ $avg\_antonyms \leftarrow$ avg_affective_value($ant$, $ac$);
8  $\quad$ $s[ac] \leftarrow$ avg($avg\_similar$, $-avg\_antonyms$);
9  **end**
10  **return** $s$;

---

### Affective Knowledge Extraction

Figure 6 provides an overview of the affective knowledge extraction method underlying the experiments discussed in Section 5.2. The component uses the expanded and contextualised affective models by transforming the input sentence into embedding space and then applying semantic reasoning to all sentence tokens. The use of Transformer language models such as BERT also considers the token's concept, i.e. disambiguating the concept prior to determining its values alongside the affective categories. In addition, dependency parsing and grammar rules provide information on the token's grammatical context, which is useful for considering negation and modifiers that determine a token's impact on the sentence's affective categories.

The affective knowledge extraction computes a feature vector based on (i) the token ($t_i$), (ii) the corresponding sentence ($s_j$) and (iii) the sentence's dependency tree ($dp_j$).

Equation 1 computes the sentence score along the affective category ($ac$) by first obtaining the $embedding_i$ based on token ($t_i$) and the context information available in sentence ($s_j$) from the chosen language model ($lm$). The algorithm then uses a proximity search and the approach outlined in Algorithm 2 to determine the value of the affective category in the context of the sentence. Finally, we compute the score for the affective category, considering negation and modifiers based on the dependency tree ($dp_j$) with the factors $n(dp_j, i)$ and $m(dp_j, i)$ respectively, as shown in Equation 2.

$$embedding_i = lm(t_i, s_j) \tag{1}$$

$$score(s_j, ac) = \sum_{i=1}^{n} m(dp_j, i) \cdot n(dp_j, i) \cdot \tag{2}$$
$$score(embedding_i, ac)$$

All algorithms were run using an in-house tokeniser to seamlessly integrate the presented approach with other components in our Web intelligence platform (e.g. text clean-up, keyword and topic extraction [59], dependency parsing [60] and NEL [57]). For classic embeddings (e.g. word2vec, GloVe) we have used the gensim[3] library, whereas for the BERT and DistilBERT models we have used wrappers on top of the Spacy[4] [61] and Transformers[5] [10] libraries.

### Evaluation

The evaluation process aimed at providing both quantitative insights into the methods' performance and qualitative results that support the quantitative assessment.

The quantitative evaluation in Section 5.2 is based on the revised *Hourglass of Emotions* model. For benchmarking the affective model expansion and affective knowledge extraction components, we created a gold standard

---

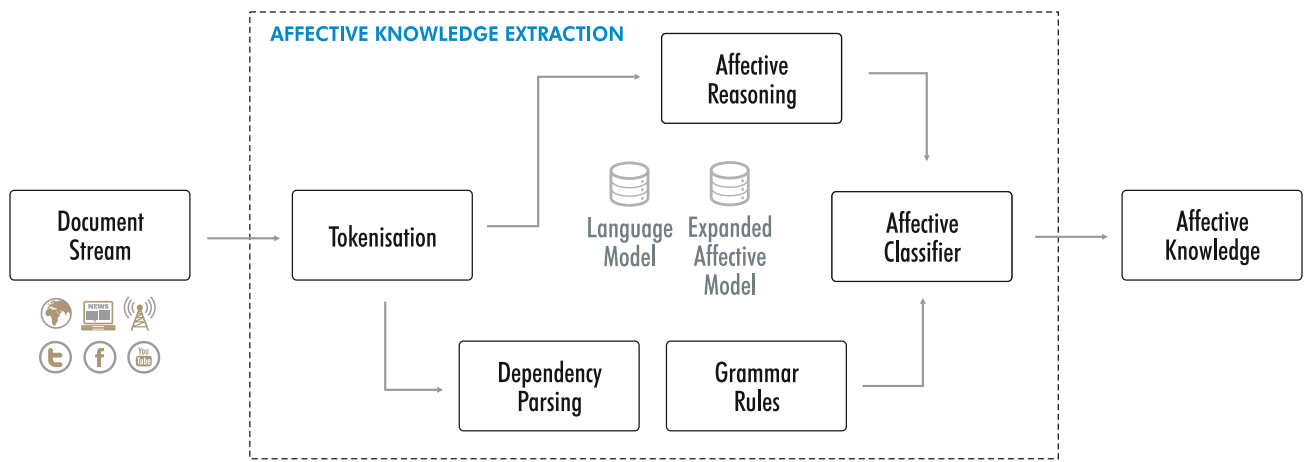

**Fig. 6** Using language models and the expanded affective models for affective knowledge extraction

comprising 346 sentences, annotated according to the affective categories defined in the model (see Section 5.1).

The qualitative evaluation focuses on domain-specific affective models used as the basis for computing the WYSDOM communication success metric. The selected model stems from the European Horizon 2020 Research Project ReTV, *Re-Inventing TV for the Digital Age*,[6] which develops knowledge extraction and visualisation services for broadcasters and media archives. The experiments illustrate how the custom model has gained in extent and consistency due to the application of the method introduced in Section 4.

### Revisited Hourglass of Emotions Gold Standard

The annotation process started with a corpus of over 12,000 sentences extracted from Wikinews. To ensure that the annotators had formed a good understanding of the revised Hourglass of Emotions model, we created a set of annotation rules and selected 500 sentences for annotation, separating some of them for testing purposes and using the rest for the gold standard. The annotation rules were collected in the *annotation guideline*, which followed rules similar to sentic evaluation challenges (e.g. SemEval [62], SMM4H [63] or WASSA

[64]). For each class (*introspection, temper, pleasantness* and *eagerness*), a set of rules was provided to guide the annotators.

The annotators were first asked to read the guidelines and to comment on them. After agreeing on the rules and their interpretation, the human experts annotated the documents using the (i) annotation guideline and (ii) tables explaining the revisited Hourglass of Emotions categories from [12]. Providing these tables improved the quality and consistency of the classifications considerably, as it helped avoiding incorrect classifications of emotions as *None* or *Unknown*. In addition, the annotation guideline contained about 50 triggers for each polar opposite of the affective categories. Selected from the dictionaries, the triggers provided cues for the annotators (e.g. a set of triggers like *awe, force, malady, defeat, terror, danger, flood, violence* that point the annotators towards emotions like *anxiety, fear* and *terror* representing the negative polarity for sensitivity). A set of examples of each polar opposite of an affective category was also selected from previously published corpora including Saravia et al. [65] and Poria et al. [66] as well as the extracted Wikinews sentences. The list of triggers and examples of emotional categories eventually helped to distinguish subtle nuances.

**Table 1** Selected examples with two senses per term for the contextualisation of seed terms based on the corresponding values assigned to the affective categories (T)emper, (I)ntrospection (A)ttitude and (S)ensitivity for these terms

| term | senses | contextualised example | T | I | A | S |
|------|--------|------------------------|-----|-------|-------|-------|
| like | a similar kind | we don't want the likes of you around here | 0.00 | 0.45 | 0.00 | 0.00 |
| | wish, care, like | Would you like to come along to the movies? | 0.00 | 0.49 | 0.00 | 0.57 |
| probe | investigation | there was a congressional probe into the scandal | 0.62 | -0.56 | -0.43 | 0.62 |
| | poke into | probe an anthill | 0.00 | 0.00 | 0.00 | 0.00 |
| project | communicate vividly | He projected his feelings | 0.00 | 0.00 | -0.76 | 0.42 |
| | throw, send | project a missile | 0.00 | -0.44 | 0.00 | -0.56 |

---

[6] www.retv-project.eu

The human annotators originated from different cultures, and the gender ratio was balanced. Selected documents were annotated twice. Annotators evaluated 120 sentences each, providing information about the sentence's affective categories (i.e. emotions that occur in the sentence), the dominant emotion and the overall polarity of the sentence. In the case of ambiguous examples, annotators highlighted the sentence and provided their reasoning behind the choice. *Unknown* was assigned to cases where it was not possible to classify the dominant emotion, and *None* to sentences not containing any emotional expressions.

Finally, an expert with relevant experience in sentic computing validated the gold standard annotations. The final gold standard was created only after a consensus between the expert and annotators had been found for difficult cases.[7] The dominant emotion was determined by the expert based on the individual annotator assessments, whereas the affective categories and polarity were averaged. Around 70% of the selected sentences were included in the final corpus (346 sentences out of 500 initially selected), with high inter-rater agreement (Fleiss kappa=0.868). *Anger* (the negative polarity of the temper category) and the category *None* turned out to be the major sources of disagreement. Although the remaining 30% of the sentences with no associated affective category were not used in the evaluation, they were kept in the corpus. Disagreements related to *Anger* were often related to cultural differences in annotator perceptions. Table 2 summarises annotator agreement across classes (e.g. the four affective categories of the revised Hourglass of Emotions model split by polarity plus the *None* class). The fact that no sentences were marked as *Unknown* supports the assumption that the revised Hourglass of Emotion model is well-suited for classifying emotions [12].

**Table 2** Agreement (Fleiss kappa) within the gold standard for positive and negative values of the affective categories (*A)titude*, (*I)ntrospection*, (*S)ensitivity* and (*T)emper* used in the revisited Hourglass of Emotions model. The *None* category indicates sentences with no emotion

| category | | agreement |
|---|---|---|
| A+ | pleasantness | 0.90 |
| A- | disgust | 0.86 |
| I+ | joy | 0.90 |
| I- | sadness | 0.85 |
| S+ | eagerness | 0.88 |
| S- | fear | 0.97 |
| T+ | calmness | 0.85 |
| T- | anger | 0.78 |
| None | | 0.78 |

## Quantitative Analysis Based on the Revisited Hourglass of Emotions Model

The presented evaluations use the F1 score, which is defined as the harmonic mean of precision and recall. The experiments performed in this section compute the evaluation scores based on two approaches:

1. The performance metric based on the *dominant affective category* determines whether the dominant emotion corresponds to the one presented in the gold standard that yields a True Positive (TP). Otherwise, the sentence is considered a False Positive (FP). The obtained score equates to the recall for the dominant emotion.
2. The second metric determines whether *all affective categories* present in a sentence have been correctly detected. If the affective category in the gold standard matches the one computed by the system, we obtain a

**Table 3** Recall of the *dominant emotion* based on the affective values encoded in SenticNet 5 and extensions: *plain* indicates results based on the application of the SenticNet 5 lexicon as a static lookup table, *+AR* signifies affective reasoning, *+L* lemmatisation, and *+GR* grammar rules

| category | | plain | +AR | +AR+L | +GR | +AR+GR | +AR+L+GR |
|---|---|---|---|---|---|---|---|
| T+ | calmness | 0.50 | 0.64 | 0.61 | 0.50 | 0.68 | 0.64 |
| T- | anger | 0.22 | 0.39 | 0.47 | 0.24 | 0.47 | 0.56 |
| I+ | joy | 0.63 | 0.66 | 0.80 | 0.60 | 0.69 | 0.80 |
| I- | sadness | 0.37 | 0.57 | 0.57 | 0.43 | 0.61 | 0.61 |
| A+ | pleasantness | 0.82 | 0.80 | 0.84 | 0.76 | 0.78 | 0.80 |
| A- | disgust | 0.49 | 0.61 | 0.50 | 0.47 | 0.58 | 0.54 |
| S+ | eagerness | 0.46 | 0.62 | 0.54 | 0.56 | 0.64 | 0.56 |
| S- | fear | 0.50 | 0.67 | 0.57 | 0.47 | 0.63 | 0.57 |
| overall | | 0.50 | 0.62 | 0.60 | 0.49 | 0.63 | 0.63 |

---

**Table 4** Precision/recall/F1 from direct application of the GloVe model on SenticNet 5. Note that despite the already large size of the lexicons, applying affective reasoning (+AR) significantly improves the results

| category | | plain | +AR | +AR+L | +GR | +AR+GR | +AR+L+GR |
|---|---|---|---|---|---|---|---|
| T+ | calmness | 0.31/0.55/0.40 | 0.34/0.58/0.43 | 0.31/0.5/0.38 | 0.29/0.50/0.30 | 0.38/0.63/0.48 | 0.34/0.53/0.42 |
| T- | anger | 0.58/0.26/0.36 | 0.64/0.40/0.50 | 0.63/0.4/0.49 | 0.54/0.26/0.30 | 0.70/0.46/0.55 | 0.67/0.47/0.55 |
| I+ | joy | 0.55/0.74/0.63 | 0.62/0.77/0.69 | 0.62/0.85/0.71 | 0.56/0.74/0.64 | 0.64/0.80/0.71 | 0.63/0.85/0.72 |
| I- | sadness | 0.69/0.45/0.55 | 0.75/0.63/0.69 | 0.79/0.58/0.67 | 0.71/0.49/0.58 | 0.78/0.64/0.71 | 0.80/0.60/0.69 |
| A+ | pleasantness | 0.53/0.75/0.63 | 0.61/0.78/0.68 | 0.57/0.83/0.67 | 0.52/0.71/0.60 | 0.61/0.79/0.69 | 0.58/0.79/0.67 |
| A- | disgust | 0.71/0.41/0.52 | 0.73/0.54/0.62 | 0.75/0.46/0.57 | 0.66/0.41/0.51 | 0.73/0.54/0.62 | 0.73/0.51/0.61 |
| S+ | eagerness | 0.48/0.46/0.47 | 0.60/0.57/0.58 | 0.56/0.52/0.54 | 0.47/0.43/0.45 | 0.60/0.59/0.59 | 0.58/0.54/0.56 |
| S- | fear | 0.59/0.53/0.56 | 0.67/0.69/0.68 | 0.62/0.62/0.62 | 0.58/0.53/0.55 | 0.67/0.67/0.67 | 0.64/0.64/0.64 |
| overall | | 0.58/0.50/0.54 | 0.64/0.62/0.63 | 0.63/0.58/0.61 | 0.57/0.50/0.53 | 0.66/0.64/0.65 | 0.64/0.61/0.63 |

True Positive (TP). For affective categories that have not been detected, the metric yields False Negatives (FN). False Positives (FP) are returned for computed categories that are not present in the gold standard. The second metric corresponds to the F1 score, listed together with precision and recall.

The first evaluations aim at quantifying the improvements from affective model expansion, based on the revisited Hourglass of Emotions as seed model. In contrast to SenticNet 5, which covers 100,000 commonsense concepts and had already been published at the time our experiments have been conducted, SenticNet 6 has only become available after June 2020 [21]. Therefore, the extension of an affective model based on the revisited Hourglass of Emotions model [12] has been the perfect candidate for testing our approach:

– It describes a sophisticated affective model that can be used in conjunction with the gold standard developed in Section 5.1 to design reproducible experiments for quantifying the impact of the expansion process on model performance, and
– Stakeholders can benefit from the expanded model since it allows extracting affective knowledge on

emotions based on the categories defined in the revisited Hourglass of Emotions model.

### Evaluation Based on SenticNet 5

The first set of evaluations based on SenticNet 5 do not yet consider the improvements of the SenticNet 6 model [21] nor extensions to this model based on the method introduced in Section 4.1. The SenticNet 5 lexicon covers many n-grams that do not necessarily have precomputed vectors in the GloVe model and are thus not annotated. This still leaves several tens of thousands of single-token terms with non-zero values for each affective dimension.

Tables 3 and 4 summarise the outcome of these experiments. The performance of a similarity-based approach that considers affective reasoning (+AR) is compared against a static lookup based on the same initial term–value lexicon. The similarity calculation is based on the *glove-wiki-gigaword-300* pre-trained model. The results indicate that the suggested expansion considerably improves the performance across all SenticNet categories. The first column (plain) describes the evaluation outcome based on the unmodified SenticNet 5 lexicon, the second column (+AR) provides the results after the model expansion, and the final column (+AR+L) shows the expanded model based on lemmas rather than the unmodified terms. All three experiments use the SenticNet 5 lexicon as the basis, and they only differ in the additional logic applied during the annotation phase. While the inclusion of grammar parsing (+GR) notably improves the result, this is not the case for lemmatisation.

**Table 5** Recall of the *dominant emotion* using the GloVe language model on SenticNet 5, with and without applying dependency parsing and grammar rules (GR)

| category | | plain | +AR | +AR+L | +GR | +AR+GR | +AR+L+GR |
|---|---|---|---|---|---|---|---|
| T+ | calmness | 0.00 | 0.39 | 0.39 | 0.00 | 0.61 | 0.61 |
| T- | anger | 0.06 | 0.71 | 0.71 | 0.06 | 0.78 | 0.75 |
| I+ | joy | 0.17 | 0.66 | 0.63 | 0.17 | 0.66 | 0.63 |
| I- | sadness | 0.04 | 0.72 | 0.67 | 0.07 | 0.80 | 0.76 |
| A+ | pleasantness | 0.07 | 0.73 | 0.75 | 0.07 | 0.75 | 0.71 |
| A- | disgust | 0.06 | 0.72 | 0.72 | 0.07 | 0.72 | 0.71 |
| S+ | eagerness | 0.05 | 0.59 | 0.59 | 0.05 | 0.62 | 0.62 |
| S- | fear | 0.07 | 0.73 | 0.67 | 0.07 | 0.80 | 0.67 |
| overall | | 0.06 | 0.67 | 0.66 | 0.07 | 0.72 | 0.69 |

**Table 6** Precision/recall/F1 for a model based on the revisited Hourglass of Emotions with simple word embeddings (GloVe): *plain* uses the small generated dictionaries for static lookup, +AR indicates affective reasoning for matching novel terms, +L lemmatisation, and +GR application of grammar rules/negation/dependency parsing

| category | | plain | +AR | +AR+L | +GR | +AR+GR | +AR+L+GR |
|---|---|---|---|---|---|---|---|
| T+ | calmness | 0.00/0.00/0.00 | 0.50/0.37/0.42 | 0.50/0.37/0.42 | 0.33/0.03/0.05 | 0.72/0.61/0.66 | 0.68/0.61/0.64 |
| T- | anger | 0.90/0.13/0.22 | 0.70/0.79/0.75 | 0.70/0.76/0.73 | 1.00/0.13/0.22 | 0.80/0.86/0.83 | 0.79/0.81/0.80 |
| S+ | eagerness | 0.02/1.00/0.04 | 0.65/0.57/0.60 | 0.62/0.57/0.59 | 1.00/0.07/0.12 | 0.63/0.59/0.61 | 0.63/0.59/0.61 |
| S- | fear | 0.33/0.04/0.07 | 0.67/0.75/0.71 | 0.66/0.69/0.67 | 0.33/0.04/0.07 | 0.67/0.71/0.69 | 0.67/0.69/0.68 |
| A+ | pleasantness | 0.50/0.06/0.11 | 0.72/0.66/0.69 | 0.72/0.68/0.70 | 0.56/0.06/0.12 | 0.76/0.70/0.73 | 0.74/0.68/0.71 |
| A- | disgust | 0.55/0.07/0.12 | 0.73/0.77/0.75 | 0.72/0.74/0.73 | 0.54/0.08/0.14 | 0.75/0.80/0.77 | 0.73/0.77/0.75 |
| I+ | joy | 0.64/0.11/0.18 | 0.73/0.71/0.72 | 0.70/0.69/0.70 | 0.78/0.11/0.19 | 0.76/0.74/0.75 | 0.73/0.72/0.73 |
| I- | sadness | 0.60/0.04/0.08 | 0.50/0.77/0.76 | 0.74/0.74/0.74 | 0.71/0.07/0.13 | 0.77/0.79/0.78 | 0.77/0.77/0.77 |
| overall | | 0.58/0.07/0.12 | 0.68/0.67/0.68 | 0.68/0.66/0.67 | 0.65/0.07/0.13 | 0.74/0.73/0.73 | 0.72/0.71/0.71 |

Since annotators were instructed to annotate the gold standard sentences with the affective categories of the revisited Hourglass of Emotions model in mind, somewhat lower results are to be expected. The evaluation used the following mapping between the affective categories in the revisited model (left) and the ones in SenticNet 5 (right) with their respective poles:

- sensitivity (*eagerness/fear*): sensitivity (*anger/fear*)
- attitude (*pleasantness/disgust*): aptitude (*trust/disgust*)
- introspection (*joy/sadness*): pleasantness (*joy/sadness*)
- temper (*calmness/anger*): attention (*anticipation/surprise*)

The mapping from temper to attention has proven to be the most problematic one yielding the lowest metrics in Tables 3 and 4.

**Evaluation Based on the Hourglass of Emotions**

The second experiment drew upon the affective categories used in the revisited Hourglass of Emotions (*temper*, *introspection*, *attitude* and *sensitivity*) and example concepts taken from [12]. For affective categories already present in the original SenticNet model, we selected the top 20 terms, ignoring illnesses, from SenticNet 5. For new categories we manually added additional terms to provide a more balanced seed set. Tables 5 and 6 illustrate the impact of the affective model expansion process on the model's performance, which considerably improves recall for all affective categories.

The first interesting observation is model performance before the expansion, which is considerably lower when compared to SenticNet 5. This was expected given the limited size of the seed model. It only covers 445 affective concepts in total, as compared to 300-500 affective concepts per affective category in the expanded lexicons. Consequently, many sentences are assigned a neutral value and most non-neutral sentences are affected by a single trigger. SenticNet 5 has a considerably higher coverage although its affective categories differ from the revised model. This is illustrated in model performance after the expansion process, which yields an almost five-fold improvement, even more than in the experiment described in Section 5.2.1. The inclusion of grammar rules (GR) further improves the results.

**Table 7** Recall of the *dominant emotion* for a model based on the revisited Hourglass of Emotions using the BERT/DistilBERT language model, with and without applying dependency parsing and grammar rules (GR)

| category | | BERT | DistilBERT | BERT+GR | DistilBERT+GR |
|---|---|---|---|---|---|
| T+ | calmness | 0.62 | 0.46 | 0.75 | 0.68 |
| T- | anger | 0.55 | 0.65 | 0.45 | 0.65 |
| I+ | joy | 0.37 | 0.46 | 0.40 | 0.43 |
| I- | sadness | 0.76 | 0.80 | 0.74 | 0.83 |
| A+ | pleasantness | 0.62 | 0.64 | 0.65 | 0.67 |
| A- | disgust | 0.68 | 0.69 | 0.69 | 0.68 |
| S+ | eagerness | 0.38 | 0.36 | 0.46 | 0.36 |
| S- | fear | 0.90 | 0.87 | 0.80 | 0.70 |
| overall | | 0.61 | 0.63 | 0.62 | 0.64 |

**Table 8** Precision/recall/F1 values for *all emotions* within a sentence using the BERT/DistilBERT language model with and without applying dependency parsing and grammar rules (GR) on a model that has been based on the revisited Hourglass of Emotions

| category | | BERT | DistilBERT | BERT+GR | DistilBERT+GR |
|---|---|---|---|---|---|
| T+ | calmness | 0.45/0.61/0.52 | 0.50/0.39/0.44 | 0.52/0.76/0.62 | 0.71/0.63/0.67 |
| T- | anger | 0.75/0.57/0.65 | 0.72/0.71/0.71 | 0.87/0.57/0.69 | 0.84/0.74/0.79 |
| I+ | joy | 0.66/0.35/0.46 | 0.73/0.42/0.53 | 0.65/0.43/0.52 | 0.71/0.38/0.50 |
| I- | sadness | 0.60/0.79/0.69 | 0.65/0.84/0.73 | 0.63/0.77/0.69 | 0.66/0.81/0.73 |
| A+ | pleasantness | 0.62/0.55/0.58 | 0.66/0.61/0.64 | 0.65/0.60/0.62 | 0.69/0.58/0.63 |
| A- | disgust | 0.65/0.64/0.65 | 0.71/0.68/0.69 | 0.69/0.64/0.67 | 0.73/0.70/0.72 |
| S+ | eagerness | 0.71/0.43/0.54 | 0.69/0.39/0.50 | 0.68/0.50/0.58 | 0.63/0.39/0.47 |
| S- | fear | 0.64/0.85/0.73 | 0.64/0.85/0.73 | 0.67/0.82/0.74 | 0.61/0.76/0.68 |
| overall | | 0.64/0.58/0.61 | 0.67/0.60/0.63 | 0.70/0.61/0.65 | 0.68/0.62/0.65 |

Tables 7 and 8 outline performance gains achieved by applying Transformer-based language models such as BERT and DistilBERT. Several other models were also tested (e.g. RoBERTa, XLNet) but since the classic BERT model (*bert-base-uncased*) and the distilled model (*distillbert-base-uncased*) yielded the best initial results, we have only considered these in the final evaluations.

## Qualitative Evaluation Based on a Domain-Specific Affective Model

As part of the requirements elicitation process in the ReTV project, domain experts from the participating media organisations were asked to provide short lists of desired and undesired associations for their organisations. Their input was condensed into the ten undesired and ten desired concepts listed in Table 9 (column *seed model*), which were then fed into the affective model expansion process. This process yielded the additional concepts listed in the column *expanded model*.

As in the experiments from the previous section, the expanded model considerably improved recall. A qualitative analysis performed on Wikinews articles showed that the suggested model extensions allowed the identification of additional affective content that would have been classified as neutral with the seed model. Table 10 lists example sentences, their corresponding desirability score and affective concepts that have been identified based on the extended model.

## Discussion

The experiments were designed to establish a baseline for the proposed method's performance on real-world affective models. The method can be used in conjunction with a wide variety of embeddings, from classic approaches such as word2vec and GloVe to more recent embeddings extracted from Transformer language models like BERT. Pre-trained and unmodified versions of GloVe, BERT and DistilBERT yielded significant improvements of the evaluated models. This is encouraging given that publicly available models were used without further optimisations. The suggested approach works well for models of varying complexity and in settings where lexical resources are scarce, as has been the case for the revisited Hourglass of Emotions model.

Results could be further improved by using custom embeddings or fine-tuning the selected language model. In such an optimised setting, BERT and DistilBERT are likely to outperform GloVe, which yielded the best results for the pre-trained models (Table 6). BERT was not necessarily built to solve all classes of NLP problems. While it has shown good results for tasks like entity recognition or basic sentiment analysis, serious issues have surfaced during evaluations of fine-grained inference problems. As shown by Ettinger [67], BERT had shortcomings in regard to the impact of negation within larger contexts. This was the main reason to develop models for improving

**Table 9** Selected concepts from the seed model (left) and the expanded affective model (right) for the broadcasting domain

| seed model | | expanded model | |
|---|---|---|---|
| desired | undesired | desired | undesired |
| balanced | boring | articulate | callous |
| captivating | censored | businesslike | convoluted |
| entertaining | disrespectful | captivating | degrading |
| informative | fake | competent | demeaning |
| innovative | irrelevant | dependable | groundless |
| investigative | offensive | dispassionate | intolerant |
| professional | partisan | enlightening | misguided |
| reliable | self-referential | entertaining | pointless |
| transparent | unbalanced | insightful | slanderous |
| trustworthy | unprofessional | unbiased | undignified |

**Table 10** Positive (desired) and negative (undesired) results obtained with the expanded affective model

| title | desirability | affective concepts |
|---|---|---|
| "Just to make this perfectly clear, I was laughing at the joke and not at any group of people." | 1.0 | just: 1.0; make: 1.0; perfectly clear: 1.0; laughing: -0.05; group: -0.05 |
| So while coverage for Democrats overall was a bit more positive, that was almost all due to extremely favorable coverage for Obama. | 1.0 | coverage: 0.05; overall: 0.2; bit: 0.05; positive:1.0; negative: -1.0; extremely favorable: 1.0 |
| The statement drew criticism to the network for being false. | -1.0 | criticism: -0.55; false: -0.7 |
| Jeet Heer, the national affairs correspondent at The Nation said "the big loser of the night was the network that hosted the event." | -0.715 | affairs: -0.05; loser: -1; Nation said: -0.05 |

BERT's negation handling, like NegBERT [68], as negation detection is important not only for clinical text analysis, but also for bias, sarcasm or the general task of affective inference. Since we are not using NegBERT-based models, our approach complements this work and can be applied on top of existing BERT models constrained by negation issues.

Table 8 sheds light on another interesting aspect: results for positive examples trail those for negative examples, regardless of method and model. This outcome suggests a negative bias within the Wikinews gold standard, which is confirmed by the literature noting that most political articles have a negative connotation [69]. Dependency parsing and rule-based syntactic processing can yield improved results, even for sophisticated language models that extract contextualised embeddings for documents in the pipeline.

Table 9 illustrates that the proposed method can help increase the explainability of affective models, since it is possible to inspect and modify the seed models and the expanded affective models, whereas Table 10 shows some positive and negative results obtained with the expanded affective model.

Future research could focus on better understanding the linguistic information encoded in the resulting models, e.g. by using structural probing [70], structured perceptron parsers [71] or visualisations—as demonstrated through BERT embeddings and attention layers visualisations like those from [30] and [72].

## Outlook and Conclusion

Advanced Web intelligence applications should be able to provide domain-specific communication success metrics tailored to an organisation's evolving communication goals. Such metrics depend on affective models that measure and classify how a brand, product or service is perceived across digital channels.

This article introduces a novel method for improving the coverage and consistency of such affective models that combines the advantages of word embeddings with the robustness of lexical approaches and knowledge graphs. It provides a flexible, fast and inexpensive method to create and expand affective models. The expressiveness of the method can be controlled by the complexity of the chosen embeddings. A quantitative evaluation confirmed that the expanded models clearly outperform the original models, even in resource-scarce scenarios. To conduct this evaluation, we have created a gold standard of Wikinews sentences that was annotated with the affective categories defined in the revised 2020 version of the *Hourglass of Emotions* model [12].

Integrating the results into WYSDOM combines the strength of established and extensively evaluated affective models with the ability to create domain-specific models on the fly, based on a limited set of desired and undesired associations that can be elicited in a single workshop with the communication experts responsible for a brand or product.

Future work to further advance this approach will focus on: (i) improving the affective reasoning components by incorporating techniques to cluster related word senses rather than using fixed, empirically chosen threshold values, (ii) incorporating additional common and commonsense knowledge sources into the expansion process, and (iii) adapting the language models to the affective model's domain.

**Acknowledgements** The authors would like to thank Katinka Böhm, Adriana Bassani and Lenka Kilian for their help in the specification of emotional categories and their participation in the quantitative evaluation.

**Funding Information** Open Access funding provided by the University of Applied Sciences of the Grisons (www.fhgr.ch). This research has been partially funded through the following projects: the MedMon project (www.fhgr.ch/medmon) funded by Innosuisse (No. 25587.2 PFES-ES); the ReTV project (www.retv-project.eu) funded by the European Union's Horizon 2020 Research and Innovation Programme (No. 780656) and the EPOCH project (www.epoch-project.eu) funded by the Austrian Federal Ministry for Climate Action, Environment, Energy, Mobility and Technology (BMK) via the ICT of the Future Program (GA No. 867551).

## Compliance with ethical standards

**Conflicts of interest** The authors declare that they have no conflict of interest and that informed consent was obtained from all evaluation participants.